\documentclass[11pt,oneside,letter]{article}
\usepackage{amssymb}
\usepackage{dsfont}
\usepackage{mathrsfs}

\usepackage[square,comma,numbers,sort&compress]{natbib}
\newlength{\fnwidth}
\newcommand{\qed}{\nobreak \ifvmode \relax \else
      \ifdim\lastskip<1.5em \hskip-\lastskip
      \hskip1.5em plus0em minus0.5em \fi \nobreak
      \vrule height0.75em width0.5em depth0.25em\fi}

\def\bqt{\begin{quotation}}   \def\eqt{\end{quotation}}
\def\beq{\begin{equation}}   \def\eeq{\end{equation}}
\def\bea{\begin{eqnarray}}  \def\eea{\end{eqnarray}}

\def\noi{\noindent}
\newcounter{saveeqn}%

\def\blist{\begin{list}{$\bullet$}{\setlength{\leftmargin}{.15in}
      \setlength{\itemsep}{0pt} \setlength{\topsep}{0pt}
      \setlength{\parsep}{0pt}}}

\def\W{$\mathscr{W}$}
\def\mW{\mathscr{W}}
\def\T{$\mathcal{T}$}
\def\mT{\mathcal{T}}
\def\R{$\mathscr{R}$}
\def\mR{\mathscr{R}}
\def\O{$\mathcal{O}$}
\def\mO{\mathcal{O}}

\newcommand{\ket}[1]{|#1\rangle}

\def\spacing{1}
\def\bls{\renewcommand{\baselinestretch}}
\bls{\spacing}

\begin{document}
\pagestyle{myheadings}
\thispagestyle{empty}
\renewcommand{\thefootnote}{\fnsymbol{footnote}}

\begin{center}
  {\Large \textbf{Reality --- an emerging representation of the world}}\\
  \small\bls{1}\normalsize \vspace{.25in} \textsc{Martin A.
    Green}\footnote{\,Electronic address:
    \ mgreen@perimeterinstitute.ca\vspace{1ex}}\\
  \vspace{1ex}
  \small\emph{Perimeter Institute for Theoretical Physics,\\
    31 Caroline St.~N., Waterloo, Ontario, N2L 2Y5, Canada}\\
\end{center}
\renewcommand{\thefootnote}{\arabic{footnote}}
\addtocounter{footnote}{-1}
\vspace{1ex}

\small\bls{1}\normalsize
\begin{center}\small
  \parbox{5in}{\noindent Some unique source --- the world,\/ \W\ ---
    must underlie all the information realized in the universe
    throughout time.  Perceived reality,\/ \R, is a progressively
    emerging representation of\/ \W\ in the form of the geometrical
    universe.  Time corresponds to the process of emergence.  When
    first represented in\/ \R, information about\/ \W\ is expressed in
    a non-localized, quantum manner.  As the emergence proceeds, most
    information becomes inaccessible (entropy), supporting the robust,
    redundant encoding of accessible records.  The past is encoded in
    and inferred from present records; the anticipated future will
    preserve present information and reveal unpredictable new
    information about\/ \W.  Emergence of the future demands
    non-unitary reduction of quantum states and increased Kolmogorov
    complexity of the quasi-classical records in terms of which the
    quantum states are known.  Given the limited information content
    of records, the quasi-classical universe lacks fine details;
    whereas the future must be uncertain to admit new information.}
\end{center}

\bls{\spacing}
\normalsize
\medskip
\section{{\sc Introduction}}\label{S:Intro}
What is existence?  Descartes asserted \emph{``Cogito, ergo sum''},
linking existence to both thought and the thinker.  Since thought
implies formation of new memories and corresponding changes of mental
state, which we perceive as existing at sequential times, Descartes's
assertion implicitly assumes time.  We can neither think nor sensibly
contemplate existence without an \emph{a priori} concept of time; time
is part of our nature as thinking entities.

The kernel of thought is information.  When we think, we acquire,
organize, remember and use information.  We attribute a cause or
source to this information, deeming it to be \emph{about} something,
say $X$.  The information thus becomes the essence of $X$; serves to
define $X$.  A maximal set of information about $X$ can sensibly be
identified with $X$.  If information about $Y$ is equivalent to a
subset of information about $X$, then $Y$ serves to \emph{represent}
$X$.  If the representation is faithful, then $Y \equiv X$.

We categorize something as \emph{physical} if and only if we judge
that our information about the thing was acquired through use of our
senses (i.e., through perception) and that the information could have
been different without resulting in a contradiction.  Otherwise, we
attribute the information to some timeless, non-physical thing such as
a mathematical object.  Our bodies are intuitively physical; the
melding of our body and mind enables our sensory perception of both
our body and other physical things.  We refer to all that is physical
as the \emph{universe}.  All our information about the universe serves
to define the universe.  All this information is interrelated, with
meaning derived from the relationships just as the words in a
dictionary are defined in terms of each other.

Each mental state that we experience corresponds to a state of the
universe that we identify as the momentary present.  Acquisition of
new information produces a new mental state corresponding to a new
perceived state of the universe.  Considerable information about the
former states of the universe persists, as memories, in each new
present.  However, our brains have limited capacity.  To optimize the
ongoing utility of memories, our brains organize and selectively store
information in an efficient manner, with lossy compression but ample
redundancy.  This entails construction of a conceptual model for the
universe that involves various forms of \emph{matter} distributed in
3-dimensional, geometrical \emph{space} and sensibly evolving over
time \citep{OKeefe}.  Time, space and matter are thus aspects of our
efficient model for the universe --- they aid our representation of
the evolving universe, although the representation will certainly not
be faithful.

Information corresponds to the properties and relationships of things.
It has no intrinsic properties and thus no existence independent of
the things it is about.  When we develop a model of the universe as a
composite of related things --- which our minds do as part of their
efficient memory system --- some information corresponds to
relationships of the components and other information to intrinsic
properties of the components.

Although we each perceive the universe independently, our consensus
about its properties makes us believe that the information that yields
each perceived universe has an underlying source that is independent
of us.  The combination of \emph{all} the information about the
universe that will be potentially accessible to us over time must be
attributed to a single, ultimate source.  We shall call that source
the \emph{world},\/ \W.  To be clear: all the information that defines
each momentarily perceived universe is information about the same
world.  It follows that the world must be a timeless, non-physical
thing.  We formalize this as follows:

\medskip\noi\textbf{Proposition 1 (World)}: \emph{The unique
  foundation for all information of which cognitive beings could, in
  principle, become aware about the past, present and future of the
  universe (including matter) is a definite, atemporal and enormously
  complex mathematical structure --- the \emph{world}.  The world is
  prior to and distinct from, in an ontological sense, the universe,
  time, perception, and physical concepts, objects and laws.}

\medskip\noi The thesis of this essay is summarized as follows:
Perceived reality is a progressively emerging, incomplete (or
unfaithful) representation of the world.  Progress of time corresponds
to the process of emergence of this representation.  The varied
mathematical and interpretational \emph{models} we use to characterize
perceived reality are structurally and conceptually distinct from the
world itself.  Physical laws constrain our models for perceived
reality such that the models are internally consistent and coherent,
and such that the modeled reality is a valid, yet limited,
representation of the world.  More advanced models for reality encode
information about the world with greater efficiency, scope and
fidelity.

Physics involves the invention, validation and use of models for
perceived reality, within the context of the world.  Models and
physical laws formally encode structural information about our
representation of the world, but they do not direct or constrain
either the representation or the world.  Revolutionary physics
theories --- Newtonian mechanics, relativity theory, quantum theory
(QT), inflationary cosmology --- have introduced radically different
models for reality that have each proven quite useful in spite of
their conceptual differences.  The ongoing utility of such diverse
models should make us hesitate to identify either perceived reality or
any model for reality with the world.  These proposals about the world
and its representation are made more explicit in section \ref{S:WR}.

Section \ref{S:Time} addresses the nature of time: How can the
incongruous concepts regarding time and causality of QT and general
relativity (GR) be resolved to integrate these theories?  How can the
special status of the perceived present be understood, and be
reconciled with the spacetime model of relativity theory?  Given the
symmetry of GR and QT under time reversal, what explains the low
entropy of the early universe?  More generally, why is past history
known but not the future?  These questions all relate to quantum
uncertainty, for which we shall find a natural explanation.

To be a valid representation of the world, perceived reality must be
self-consistent.  Individual views of reality must be mutually
compatible.  In section \ref{S:Univ} we consider how this consistency
and compatibility are reflected in and constrain theoretical models.
Deterministic models such as GR and QT encode and preserve
already-known information about the world.  Acquisition of new
information necessitates quantum state reduction, and the creation of
new records that are expressed through geometry and other
quasi-classical descriptors of the universe\footnote{I use
  \emph{quasi-classical} to refer to observable properties of the
  universe about which two distinct observers should reliably agree.
  This need not entail the deterministic evolution of a
  \emph{classical} universe.}.  These descriptors serve only to encode
the information --- they have no independent existence.

In section \ref{S:Topoverse}, we explore the possibility that the
world is, quite simply, a smooth 4-manifold with arbitrary global
connectivity but no inherent geometry, fields, or other decoration.
The global topology of\/ \W\ admits a representation with features
strikingly similar to a Feynman graph.  This suggests a correspondence
between the topology and the virtual particles and interactions of
quantum field theory.  Information about the topology might thus be
represented as the dynamically interacting matter and spacetime that
we perceive.

The final section provides a summary and discusses implications that
will be pursued in future work.

\section{{\sc Perceived Reality}}\label{S:WR}

Being unique and atemporal, the world does not change, nor can
temporal change be found within it.  In particular, the world should
not be mistaken for spacetime or a block universe
\citep{Ellis-2006-38}.  However, combining Proposition 1 and
the expectation that there exist \emph{representations} of\/ \W\ that
are related by their nested information about the world leads to:

\medskip\noi\textbf{Proposition 2 (Nature)}: \emph{Nature ---
  perceived reality and all its generic and specific properties
  throughout time; the subject of all information gained from
  perception --- is an unfaithful, progressively emerging
  representation\/ \R\ of\/ \W.  Time corresponds to the process of
  emergence through which, starting with minimal information at the
  Big Bang, information about\/ \W\ is cumulatively encoded in the
  evolving\/ \R.  At any given stage of emergence,\/ \R\ holds only a
  small fraction of all information about\/ \W; most details of\/ \W\
  are unknowable.}

\medskip\noi We exist, within nature, not as direct properties of\/
\W, but as players in a progressively emerging representation of\/ \W.
While we shall generally speak of\/ \R\ in the singular, it is by no
means unique.  At a minimum, each cognitive being / observer\/ \O\ has
its own variant $\mR_{\mO,t}$ at each perceived time $t$.  The process
of emergence of\/ \R, which is perceived as passage of time, preserves
(to the extent possible) that information about\/ \W\ that has already
been incorporated in\/ \R, while adapting and refining\/ \R\ to
efficiently, coherently and consistently incorporate new information.
Time serves to label the distinct representations, but also resides
within each\/ \R\ as the natural ordering of the accumulating
historical records (and memories) from which we infer prior states of
the universe.

Given Proposition 1, the generic features of both the emergence
process and the representation\/ \R\ that it yields can arise only
from the structure of\/ \W.  These features are manifested as physical
concepts, objects, and laws.

\medskip\noi\textbf{Proposition 3 (Physics)}: \emph{Physical concepts,
  objects and laws characterize and constrain \emph{models} for
  generic features of\/ \R\ such that the models are compatible with
  both cognitive perception and the generic structure of\/ \W.
  Physics arises from but does not constrain\/ \W.  Physical laws are
  conditions, within the models, under which the modeled
  representation\/ \R\ will always be coherent and internally
  consistent.}

\medskip\noi Each of the above Propositions has introduced a
corresponding conceptual layer: (1) the world, the atemporal
foundation; (2) the emergent representation of the world that is
perceived reality; and (3) mathematical and interpretational models
developed by physicists to characterize perceived reality.  Each
successive layer conveys information about the layer(s) beneath it,
but has no influence on or control over the lower layer(s).  We exist
as active participants in the middle layer.

The mathematical and interpretational frameworks of QT and GR ensure
that our layer-3 models are coherent and consistent.  The Standard
Model of particle physics serves to characterize generic aspects of
the structure of\/ \W; whereas cosmological models characterize the
global evolution of\/ \R\ while maintaining compatibility with\/ \W.
When physical laws, theories and models lead us to unreasonable
results, then either they do not provide valid characterization of
perceived reality, or we have attempted to apply them beyond their
domain of validity.  Understanding the relationship of QT and GR to
the lower layers will be an essential step towards resolution of
unreasonable results --- such as paradoxes, or issues involving
singularities --- and the satisfactory integration of these theories.

\section{{\sc Time}}\label{S:Time}

Issues of time have presented the greatest barrier to consistent
integration of GR and QT \citep{Ish4602856,Butterfield:1998dd}.  We
shall explore in this section how associating time with the
\emph{process of emergence} of the representations\/ \R\ helps
elucidate time's many aspects.  The scope and relationship of the
classical and quantum domains will become clear, and quantum
uncertainty and state reduction will be recognized as necessary in
order to allow accumulation of information.

Consider the representations $\mR_{t_i}$ corresponding to some
cognitive observer's perceived reality at her times $t_i$.  According
to Proposition 2, the advance of time corresponds to accumulation of
information about\/ \W, meaning that the information represented by
$\mR_{t_2}$ is a superset of the information represented by
$\mR_{t_1}$ whenever $t_2 > t_1$.\footnote{There may also be many
  representations of\/ \W\ that do not involve such accumulation of
  information, but they will not concern us.}  Gaining new information
implies a change in mental state, which is attributed to advancing
time.  The nesting relationship naturally determines the direction of
time.

We each exist within a personal representation\/ \R\ that includes
ourselves and our memories --- we always perceive this as the present.
The nested sub-representations of\/ \R, reflected in the organization
of our memories, express the history of the universe that has led to
our present state.  Although it derives from present memories and
physical records, we attribute this history to the notional past.  We
have no memories or records of the future, but we can anticipate the
future based on our present knowledge and the requirement that present
information be maximally preserved (although it may become
inaccessible).  Information about\/ \W\ that is not yet incorporated
into\/ \R\ must, to avoid contradiction, be truly unknown; aspects of
the future that are not mandated by compatibility with the present can
only be discovered.

Accepting no observer as special, and realizing that every present is
transient and will become part of history, we require that all
representations\/ \R, including their nested sub-representations and
future representations that can only be anticipated, must have
equivalent properties.  Of course, a dominant requirement remains that
all of these must represent different overlapping subsets of
information about the same\/ \W.  As minor players, our perception
of\/ \R\ gives us limited access to just the most persistent
information.  Due to its persistence in physical records, much of this
information can be accessed as required to update our memories.  Our
existence in each other's perceived reality mandates mutual
consistency of our shared histories.

Viewed here as a synthetic concept, the past provides a robust and
efficient mechanism for organizing and encoding our accumulated
present information about\/ \W.  This suggests a new perspective on
causality and determinism in which focus shifts from quantum states
and spacetime events to the globally represented information.
Determinism corresponds to reliable preservation of information as
time advances; causality requires future representations to
incorporate new information about\/ \W\ in a manner that also ensures
preservation of current information.  The need to respect this orderly
accumulation of information constrains our spacetime and quantum
models.

Deterministic theories, such as GR and classical electrodynamics,
preserve information as time advances.  But such theories merely
transform their dynamical model of the universe at some reference time
to represent the same information in different ways at different
times.  Deterministic theories are thus suited to modeling of the
evolving representation of already-known information in a manner that
effectively decouples it from new information.  Because they neither
add nor remove information, they fail to distinguish the present or
identify a preferred direction of time.

The unitary evolution of QT is also deterministic.  Projected future
quantum states and potential outcomes are based solely on presently
available (or hypothetical) information.  But QT, by predicting only
the \emph{probabilities} of uncertain outcomes, leaves an opening for
acquisition of new information.  Non-unitary quantum state reduction,
which has to many people been anathema, now reveals itself as an
essential process through which novel information about \W\ is added
to\/ \R\ while maintaining full compatibility with the past.  Being
about\/ \W, this new information has no \emph{causes} within\/ \R\ ---
it is simply discovered.  While the properties of\/ \W\ limit the
valid representations\/ \R, there remains freedom, illustrated by the
quantum Zeno effect, for cognitive acts of observers to guide the
\emph{order} in which new information is integrated into the
representation.  This freedom is manifested as \emph{free will}.

Information becomes persistent through its redundant expression in the
quasi-classical universe.  It is encoded in models as
deterministically evolving active degrees of freedom of geometry and
matter.  Information that has persisted from the earliest times is now
the most highly compressed and redundantly encoded in\/ \R, and is
thus highly accessible to diverse observers.  Compression also entails
entanglement, since given information is spread across many records.
Cosmological models account for an enormous range of observations
using just a few parameters \citep{Lahav-2006-33}.  In contrast,
information that is highly localized in the geometrical model for\/
\R\ may only be accessible to a limited set of observers and may
quickly become inaccessible as the representation evolves.

With this perspective on time and perceived reality, information takes
a leading role while spacetime fills a supporting role as a modeling
tool.  Geometrical spacetime and localized events are model-dependent
abstractions, useful because they enable efficient representation of
information.  When that information is determined to have entered the
representation in the distant past or remote locations, its encoding
as geometry and matter will be necessarily coarse-grained and
non-local.  To represent new information as time advances, our models
must become increasingly complex.  Accessible information is encoded
in the parameters of quasi-classical records.  Redundancy makes
records robust, but at the cost of making even more information
inaccessible --- as entropy.  Advancing time increases both
information and entropy.  Projecting backward, one must conclude
that\/ \R\ at earlier times had progressively less information and
entropy, with both eventually going to zero at the Big Bang.

\section{{\sc The Consistent Universe}}\label{S:Univ}

Every observer's universe, including the past they infer from present
records, must be self-consistent.  Partial models of the past, based
on distinct subsets of valid present records, must never imply
incompatible past events.  Existence of an inconsistency or
incompatibility would indicate that some present information, some
knowledge or memory, is not valid.  For observers to share the same
universe, their representations of the world must be mutually
consistent.  To the extent their histories overlap, they must have
compatible, although generally not identical, representations of the
matter content, interactions, and temporal order.

A desire to formalize this consistency and compatibility motivates the
search for a single mathematical and interpretational model that
accounts for core features of the representations\/ \R.  Physical
laws, as implied by Proposition 3, are merely constraints on this
model that ensure both self-consistency and compatibility with the
generic structure of\/ \W.  A successful model will provide an
efficient, highly compressed encoding of the information in\/ \R\
about the generic and specific structure of\/ \W.

We use our perceptions to test and update the details of this model.
Our biological heritage causes us to do much of this subconsciously
\citep{OKeefe,Ruffini07}.  Our intellectual capabilities enable us to
access presently available records, including the records of
experiments and past intellectual activity by ourselves (memories) and
others (writings, creations, etc.), and to interpret them within our
model's framework to acquire much more elusive information.
Development of a consistent model of the cosmos and its history,
supported and corroborated by creatively acquired, presently
accessible data, is a prime example of the sophisticated intellectual
approach to acquisition and efficient representation of information
about the world.

Any given model of reality should be expected to have a limited domain
of validity, even though the limits may not be obvious.  Stretching a
model beyond its domain, or simultaneous application of two
incongruous models, can produce confusing or unreasonable results ---
hence the difficulty of extending QT and GR into each other's natural
domain.  Given their undeniable success, we have no thought of
abandoning geometrical spacetime, GR or QT for alternative models.
Instead, Propositions 1 to 3 and our remarks on time will help us to
understand why these models work, what are their domains of validity,
how their consistent integration can be achieved, and how their
interpretation can be improved.

With QT and GR, we attempt to consistently model information about\/
\W, represented in\/ \R, as matter fields and quasi-classical, locally
Minkowskian geometry on (3+1)-dimensional spacetime.  However, the
geometry and matter inferred from present knowledge are necessarily
uncertain and coarse-grained --- their specific forms have meaning
only to the extent warranted by the information in\/ \R\ to which we
have access.  This accessible information is redundantly encoded in
the geometry and field configurations
\citep{zurek-2003-75,ollivier-2005-72,blumekohout-2006-73,zurek-2007},
including in the physical representation of intellectual records
(writing, etc.).  Individual observations and present quantum states
gain meaning only through reference to the reliably persistent,
quasi-classical universe.

One might be tempted to identify\/ \W\ as the global quantum state
$\ket{\Psi}$ of the cosmos, in the Heisenberg representation.  But QT
cannot then tell us how to factor $\ket{\Psi}$ into the perceived
quasi-classical environment and distinguished quantum subspaces
(particles, spins, etc.).  If we assume that QT applies universally
and contemplate arbitrarily large subspaces (incorporating apparatus,
laboratory, planet, etc.), then eventually we will reach a bound
beyond which the remaining quasi-classical records hold insufficient
information to fully determine the quantum state.  That bound will
define a limit to the applicability of quantum theory.

Only quantum state reduction allows us to accumulate information over
time.  To preserve newly acquired information, state reduction must
directly increase the Kolmogorov complexity \citep{Ming93} of the
quasi-classical universe in terms of which the reduced quantum state
is defined.  This will also make our model more fine grained.  It
follows that, contrary to conventional QT, the Hilbert space of
quantum states must steadily grow.

Unable to predict what new information will actually be discovered
about the world, QT covers all bets by giving unbiased treatment to
all possibilities consistent with the present.  But since the universe
must always be self-consistent, each hypothetical future must have
mutually consistent matter and geometry.  The uncertain future of
geometry must be held in lock-step with that of matter.

\section{{\sc The Structure of the World}}\label{S:Topoverse}

Although the world is beyond direct perception, we might still infer
its generic structure and make testable predictions that confirm or
deny the inferred structure.  As a plausible example, we consider the
following \citep{Green80}:

\medskip\noi\textbf{Conjecture (Topoverse)}: \emph{The world is a
  unique, connected, smooth 4-manifold,\/ \T, with unconstrained
  global connectivity but no inherent geometry, fields, or other
  decoration.}

\medskip\noi Although $\mT\equiv\mW$, we shall call\/ \T\ the
\emph{topoverse} to emphasize its topological nature.  Points and
local neighbourhoods in\/ \T, being all equivalent, contribute no
information to perceived reality.  However, the non-local, relational
information residing in the global structure of\/ \T\ is unbounded.
By Proposition 2, perceived reality is an emerging representation of
this global topology.

Proposals that complex 4-manifold topology is relevant to physics have
a long history, originating with consideration of quantum fluctuations
in spacetime geometry at the Planck scale.  Wheeler imagined that
exciton-like superpositions involving many different topologies could
give particles \citep{Wheeler57}.  Not satisfied with fluctuating
geometry driving topology changes, he envisioned some unknown
\emph{pregeometry} from which the geometrical universe would emerge.
References to many proposals regarding pregeometry are provided by
Gibbs \citep{Gibbs:1995gj}.  Sums over topologies are invoked in some
approaches to loop quantum gravity and spin foams \citep{Rei6889135},
but still within a geometrical framework.  This \emph{baking in} of
the geometrical character of space and/or time that will eventually
emerge is a common theme \citep{Meschini:2004si} --- true pregeometry
is elusive.

Novel features of the current proposal are: (i)~\T\ is unique, whereas
other models consider Feynman sums over all or large classes of
4-manifold topologies \citep{Hawking78,Rei6889135}; (ii)~the global
connectivity of\/ \T\ serves as the foundation for all information,
whereas other models invoke additional structure, such as geometry,
fields, group representations or causal structure, while often
ignoring the topological information; and (iii)~quantum theory does
not apply to\/ \T, but characterizes the emergent representations\/
\R\ of\/ \T.

Our conjecture, above, is motivated by the following construction of a
representation of\/ \T\ with striking, if simplistic, resemblance to a
Feynman graph: Assign an arbitrary Riemannian metric to\/ \T\ and pick
any point $x$.  Construct the 3-surfaces $\Sigma_r$ whose points have
the minimum metric distance $r$ from $x$.  Almost every $\Sigma_r$
will be a compact 3-manifold whose topology can be uniquely expressed
as a connected sum of a finite number of prime summands
\citep{Hempel76}.  (While the sphere, torus and projective sphere are
the only prime 2-manifolds, there is a countably infinite variety of
prime 3-manifolds characterized by orientability, spinoriality,
chirality and other properties \citep{Giu4727904}.)  Adjust the
geometry so that the prime objects are small compared to their
separations.  As $r$ increases, $\Sigma_r$ will pass through critical
levels at which the topology changes.  These changes are elementary
3-manifold cobordisms involving only a small number of prime summands.
The result is a graph-like structure with labeled edges (prime
summands) and unlabeled vertices (cobordisms).  The metric is no
longer needed.

This construction suggests a correspondence of prime 3-manifolds (or
equivalence classes thereof) to particle fields, and of elementary
cobordisms to field interactions --- particles and their interactions
arise from pure topology.  QT and GR provide the mathematical
framework through which the simplistic correspondence is transformed
into a time-space-matter representation of\/ \T.  The graph that is\/
\T\ corresponds to pregeometric, \emph{bare} or \emph{virtual}
particles.  A sufficiently coarse-grained view of the graph yields the
emergent geometry and field representation,\/ \R, with the
mutually-consistent properties of the \emph{dressed} or
\emph{physical} matter we perceive.  Quantum states are global and
nonlocal because they must model the representation in\/ \R\ of
information about\/ \T\ as a whole, rather than information about some
localized portion of\/ \T.  Anything not interconnected with us would
be beyond perception and thus not part of our reality.

\section{{\sc Discussion}}\label{S:Concl}

We have cast the unique, atemporal world,\/ \W, as the information
source underlying perceived reality,\/ \R.  The process of emergence
of\/ \R\ corresponds to the advance of time.  Time is also expressed
in the entangled relations of records that comprise perceived reality.
We exist within\/ \R\ --- always perceiving it as our present.  We
invent the past to enable our efficient compression of information in
mental and theoretical models for\/ \R.  Since information always
accumulates, the past must begin with zero information, zero entropy,
and no records: no spacetime, no matter.  The Big Bang thus represents
the birth of spacetime geometry --- albeit extremely coarse-grained
and uncertain --- and the first representation of information about\/
\W\ as matter.  To maintain consistency and respect entanglement,
macroscopic features of the universe must evolve deterministically
according to classical laws including GR.

All possible futures must be consistent with the present; but unlike
the Many Worlds Interpretation of quantum mechanics, most of these
hypothetical futures can never be realized due to incompatibility
with\/ \W.  Nonetheless, many futures will be compatible with\/ \W.
We can thus imagine \emph{many representations}: universes in which
information about\/ \W\ is incorporated in different orders.  Although
highly constrained by\/ \W, these representations will be guided by
the free will of their cognitive inhabitants who may formulate
theories, conduct experiments, write essays, and so on.

QT allows us to model perceived reality using a continuum spacetime
even though we have only finite information about\/ \W\ to encode in
observable quantum subspaces.  QT also allows us to assign
probabilities to the ways in which potential new information could
combine with present information to shape our future reality.  The
actual new information will depend on\/ \W.  That it is definite
should not be a surprise; the various alternatives allowed by QT are
just an expression of our unavoidable ignorance.

To the extent they gain meaning from records, quantum subspaces will
be perceived to interact with geometry.  Semi-classical coupling will
adequately describe their relation to the geometry (and other
quasi-classical descriptors) of the present and near past.  Predicting
future geometry is more problematic, since a unitarily evolved quantum
state cannot tell us which one of the many potential futures will
actually occur.  Whatever future we do perceive will have definite,
but still coarse-grained, mutually compatible geometry and
quasi-classical records.  The compatibility requirement means that
spacetime geometry cannot have quantum degrees of freedom independent
of the matter degrees of freedom; uncertainty of future geometry
arises solely from the uncertain quantum behaviour of matter.

Inspired by Wheeler's pregeometry, we have conjectured that the world
is a \mbox{4-manifold},\/ \T\ (the topoverse), with unconstrained
global connectivity but no geometry or other decoration.  A graphical
representation of\/ \T, whose edges correspond to prime 3-manifolds
(the particle fields) and vertices to elementary cobordisms (field
interactions), can be thought of as the complete Feynman graph of the
universe for all time.  Classification and characterization of the
prime 3-manifolds and determination of their elementary cobordisms
should reveal exact and approximate symmetries that can be placed into
correspondence with Standard Model particles.

The universe is our personal and collective consistent representation
of information about the world.  It remains a puzzle how the willful
acts of one observer can affect the futures of all observers.  Perhaps
our cognitive perception is also entangled, and our individuality is
an illusion.  One might then say that the universe is alive, and we
are its neurons.

\section{{\sc Acknowledgements}}

I thank Harvey Brown, Olaf Dreyer, Stephen Green, John Moffat and Hans
Westman for helpful discussions and feedback.  I am grateful to the
Perimeter Institute for hospitality.

\end{document}